\documentstyle[12pt,aaspp4]{article}
\def\deg{\hbox{$^\circ$}} 
\def\arcmin{\hbox{$^\prime$}} 
\def\arcsec{\hbox{$^{\prime\prime}$}}

\def\fs{\hbox{$.\!\!^{\rm s}$}} 
\def\fdg{\hbox{$.\!\!^\circ$}} 
\def\farcm{\hbox{$.\mkern-4mu^\prime$}} 
\def\farcs{\hbox{$.\!\!^{\prime\prime}$}} 
\def\NH{$N_{\rm HI}$}
\def\hi{H\,{\sc i~}} 
\def\hii{H\,{\sc {ii~}}} 
\def\kms{km\,s$^{-1}$}

\lefthead{Burton, Braun, Walterbos, \& Hoopes}
\righthead{Discovery of a Nearby LSB Galaxy}
\begin{document}
\title{Discovery of a Nearby Low--Surface--Brightness Spiral Galaxy}

\author{W.B. Burton}
\affil{University of Leiden Observatory, P.O. Box 9513, 2300\,RA 
Leiden, Netherlands}

\author{R. Braun}
\affil{Netherlands Foundation for Research in Astronomy, P.O. Box 
2, 7990\,AA Dwingeloo, Netherlands}

\and

\author{R.A.M. Walterbos and C.G. Hoopes}
\affil{Department of Astronomy, New Mexico State University, Las 
Cruces, NM 88003}

\begin{abstract}During the course of a search for compact, isolated gas 
clouds moving with anomalous velocities in or near our own Galaxy 
(\cite{bra98}), we have discovered, in the data of the Leiden/Dwingeloo 
survey (\cite{har97}) of Galactic hydrogen, the \hi signature of a large 
galaxy, moving at a recession velocity of 282 \kms\, with respect to our 
Galaxy.  Deep multicolor and spectroscopic optical observations show the 
presence of star formation in scattered \hii regions; radio \hi synthesis 
interferometry confirms that the galaxy is rich in \hi and has the rotation 
signature of a spiral galaxy; a submillimeter observation failed to detect 
the CO molecule.  The radio and optical evidence combined suggest its 
classification as a low--surface--brightness spiral galaxy.  It is located 
in close spatial and kinematic proximity to the galaxy NGC\,6946.  The 
newly--discovered galaxy, which we call Cepheus~1, is at a distance of 
about 6 Mpc.  It is probably to be numbered amongst the nearest few LSB 
spirals. \end{abstract}


\section{Introduction} 
     The census of even large, nearby galaxies is likely to be incomplete.  
Extinction of light due to intervening interstellar matter hides some 20\% 
of the extragalactic sky (\cite{kra98,bal94,see96,kar98}) at optical 
wavelengths, prejudicing against the discovery of all types of galaxies.  
An additional bias pertains to the class of low--surface--brightness 
galaxies, which, although carrying substantial mass, have modest stellar 
luminosities because of their slow evolution.  The bias against discovering 
LSB systems (\cite{imp97,bot97,deb97}) becomes even more severe under 
conditions of substantial extinction.  

     The Zone of Avoidance traces the accumulated optical extinction 
resulting from overlapping interstellar clouds, forming an irregular band 
along the Zodiac.  The swath of obscuration is particularly broad in the 
constellation Cepheus. In 1920, Pannekoek had painstakingly drawn the 
contours of the Milky Way from counts of stellar surface density, and had 
devoted particular attention to the Cepheus region: his contours show a 
cleft of low stellar density (not recognized yet in 1920 as due to 
obscuration by dust) rising up from the Galactic equator and crossing the 
direction of the galaxy whose discovery is reported here.  In 1934, 
Hubble had traced the distribution of some 44,000 extragalactic systems, 
and listed the region as the Cepheus Flare, which extends as far as 
$20\deg$ above the equator of the Milky Way, amongst those containing {\em 
no} galaxy.  

     The searches for hidden galaxies which rely on detecting the radio 
signals at 21--cm, specifically the Dwingeloo Obscured Galaxies Survey 
(\cite{hen98,riv98}) and the Parkes Multibeam Survey (\cite{sta98}), have 
been confined to about $5\deg$ of the Galactic equator and have 
given priority attention to directions where known cosmological structures, 
most notably the Supergalactic Plane, cross the Milky Way.  The principal 
optical and infrared searches for hidden galaxies cover larger swaths of 
the sky, but did not detect this system of low optical brightness.  The new 
galaxy is at a higher galactic latitude, $8\deg$, than covered in radio ZoA 
searches, and is located at a large angle (some $45\deg$) from the 
Supergalactic Plane defined by the galaxies in the Local Supercluster. The 
Nearby Galaxies Atlas (\cite{tul87}), comprising 2367 galaxies with 
velocities less than 3000 \kms, shows only the galaxy NGC\,6946 and its 
dwarf companion UGC\,11583, within a circle of diameter more than $30\deg$ 
encompassing the Cepheus Flare region (except for NGC\,6951 which, at a 
recession velocity of more than 1700 \kms, is so distant as to be 
irrelevant here).  The obscuration in this region is substantial, but only 
about half that of the average at latitudes within $5\deg$ of the galactic 
equator.  This so--called Local Void (\cite{tul87}) is probably 
intrinsically sparsely populated, although, as we remark further below, 
the DOGS effort has found three probable additional galaxies accompanying 
NGC 6946. 

\section{Observations}

\subsection{Discovery as an Interloper amongst \hi High--Velocity Clouds}

     Although substantially opaque at optical wavelengths, interstellar 
matter is largely transparent to radio radiation.  The new galaxy was 
discovered during an effort (\cite{bra98}) to identify and characterize a 
class of compact, isolated high--velocity clouds.  HVCs are objects 
detected in \hi emission, moving at velocities inconsistent with the 
kinematics of the Milky Way as currently understood.  Their distances 
remain largely uncertain (\cite{wak97}); there is in particular no measure 
of the distance to any of the compact, isolated objects.  The compact, 
broad \hi signature of many of these objects resembles what one also might 
expect from a gas--rich nearby galaxy.  \hi investigation of such lines can 
be complicated by radio interference:  we dealt with the possibility of 
interference contamination by using the Dwingeloo 25--meter telescope to 
observe, anew, all candidate members of the class of compact HVCs we had 
found in the Leiden/Dwingeloo survey (\cite{har97}), on a dense, 
fully--sampled grid.  This confirmed the reality of the signals 
(interference is not repeatable) and furthermore tightly constrained the 
direction of the unresolved features to an angular accuracy of about $5'$.

     One spectrum amongst those of some 60 confirmed compact HVC's drew our 
particular attention.  The spectrum shown in Fig. 1 resembles the discovery 
spectrum (\cite{kra94}) of the galaxy Dwingeloo 1, now recognized as the 
nearest, grand--design barred spiral: both show a broad, weak, \hi 
signature at such a modest velocity as to blend with foreground emission 
from the Milky Way. We noticed also that the spectral feature in question 
had a velocity inconsistent with the other compact HVCs in our list.  The 
recession velocity of the suspect feature is $v_{\rm LSR}=65$ \kms\, 
measured with respect to the Local Standard of Rest reference frame 
conventionally used in high--velocity cloud work; the corresponding 
heliocentric velocity is $v_{\rm HEL}=58$ \kms.  Correcting for the motion 
of the LSR within our Galaxy gives a velocity relative to the Galactic 
Standard of Rest of $v_{\rm GSR}=v_{\rm LSR} + 220 \cos(b)\sin(l)=282$ 
\kms.  The kinematics of the class of compact HVCs (\cite{bra98}) resembles 
that of the members of the Local Group (\cite{gre97}): but there is no 
Local Group member, and no other compact HVC, with a $v_{\rm GSR}$ as large 
as that measured for the suspect feature.  We searched the standard 
catalogs of emission and reflection nebul\ae, planetary nebul\ae, and \hii 
regions, finding no plausible alternative to an extragalactic system.  

\subsection{Optical Data}
     We extracted the field containing the suspect \hi feature from the 
STScI Digitized Sky Survey CD--ROM, providing a scan of the optical Palomar 
Observatory Sky Survey red plate.  Within the error circle determined by 
our new Dwingeloo data, we found a faint smudge, centered near 
$\alpha,\,\delta\,({\rm J2000}) = 20^{\rm h}51^{\rm 
m}10\fs6,\,+56\deg53'25\farcs0$, or $l,b = 94\fdg37,+8\fdg01$.  We then 
contacted M. Irwin, who manages the APM microdensitometer at the Royal 
Greenwich Observatory. The scans he kindly produced gave higher resolution 
than the DSS, and, more importantly, included all three colors of the POSS 
II data; they confirmed the reality of the optical counterpart to the \hi 
signal, and gave, in the color information, a discriminant against a 
Galactic reflection or emission nebula. 

     After the \hi signature had been confirmed in new Dwingeloo data, and 
the optical counterpart found on the APM scans, we secured Director's 
Discretionary Time on the ARC 3.5--meter telescope of the Apache Point 
Observatory, some Netherlands service time on the James Clerk Maxwell 
Telescope, for an attempt at a CO detection; and approval for detailed \hi 
interferometry on the DRAO seven--antenna Synthesis Telescope in Penticton, 
British Columbia. 

     We observed the region with the 3.5--m telescope at Apache Point 
Observatory to determine the nature of the optical counterpart. First, 
using the DIS camera, green (Gunn g: 5240\,\AA) and red (Gunn r: 6700\,\AA) 
images were obtained of a single $4\arcmin \times 4\arcmin$ field centered 
on the POSS II optical smudge. Subsequently, using the SPICAM camera, we 
imaged a $7\arcmin \times 14\arcmin$ region in the broadband red (R: 
6600\,\AA) and near--infrared (I: 8000\,\AA) bands, using the 
drift--scanning mode to obtain a larger field of view. The effective 
exposure time was 10 minutes in both filters; the image resolution was 
typically between $1\farcs2$ and $1\farcs5$. We also observed a single 
$4\farcm4 \times 14\arcmin$ strip for 5 minutes in a narrow (25\,\AA\, 
passband) filter centered at 6561\,\AA, close to the rest wavelength of the 
H$\alpha$ line. This allowed a search for ionized nebul\ae\,created by 
massive hot stars in the velocity range $-600 < v_{\rm LSR} < +500$\,\kms.  
We removed continuum emission from stars in the narrow filter by 
subtracting a suitably--scaled R--band image. Some 30 \hii regions were 
detected in this short exposure, distributed over a spatial extent of 
$5\farcm5$ diameter. Figure 2 shows a color composite of the optical data.  

     The detection of \hii regions was important because optical 
spectroscopy of \hii regions is a powerful tool to further elucidate the 
nature of Cepheus~1. We obtained spectra of two \hii  regions close to the 
optical center of the galaxy with the DIS spectrograph on the 3.5--m 
telescope, covering wavelengths from 4600 to 5300\,\AA, and from 5700 to 
6800\,\AA. The radial velocities of the \hii regions are $v_{\rm LSR}=50\pm 
5$ \kms\, and $v_{\rm LSR}=58\pm 5$ \kms, respectively. We derived the 
total optical extinction towards the \hii regions from the flux ratio of 
the H$\alpha$ to H$\beta$ recombination lines. The Balmer--decrement values 
are A$_{\rm V}=2.45$ and A$_{\rm V}=2.78$, respectively, in good agreement 
with the Galactic foreground extinction of A$_{\rm V}=2.0$ deduced from the 
Galactic \hi column density, because \hii regions typically have 0.5 to 1.0 
magnitudes of extinction due to dust in the galaxy in which they are 
embedded. Taking this internal extinction into account, we adopt A$_{\rm 
V}=2.0$, consistent with the foreground \hi column and the corrected Balmer 
decrements.  (We note that extinction extrapolated from the DIRBE dust 
analysis by Schlegel, Finkbeiner, and Davis (1998) yields A$_{\rm V}=3.1$; 
our determination of the extinction using the Balmer decrement is a direct 
one, and refers to the exact direction of the galaxy, and thus is to be 
preferred over the extrapolated, relatively low resolution, DIRBE result.) 

     We determined the optical morphology and integrated stellar luminosity 
of Cepheus~1 from the broadband images, after correcting for Galactic 
foreground extinction.  We summed the optical light in elliptical 
apertures, after removing foreground stars from the images. The light 
profile can be described by a single exponential with a scale length 
of $60\arcsec$ (1.8 kpc for the distance of 6 Mpc defended below), but 
there is considerable uncertainty in the exact shape because of the 
faintness of the emission and the large number of foreground stars.  The 
best fitting ellipses are strongly flattened (axial ratio about 0.3 to 
0.5), which seems in contradiction with the much more circular appearance 
of the \hi and \hii region distributions.  It is possible that the 
continuum light we detect is mainly contributed by a faint stellar bulge or 
bar, while the actual disk may be mostly too faint to be seen at all; the 
\hii regions extend well beyond the radius to which we can trace the 
continuum light.  However, if we do attribute the light detected here to a 
disk, we derive the following parameters.  The central disk surface 
brightness, corrected for foreground extinction, is 21.9 mag\,arcsec$^{-2}$ 
in R, and 22.8 to 23.0 mag\,arcsec$^{-2}$ in B. (Galaxies with such central 
surface brightness would be selected against, even under moderate 
foreground extinction.)  Cepheus 1 has a faint stellar disk which, in the 
current data, can only be traced over an angular diameter of $3\arcmin$, 
about half the extent of the \hii region distribution. Converting the 
de--reddened g--r color, derived from the calibrated DIS images, to 
standard Johnson passbands following Jorgensen (1994), we find B--V=0.57, 
typical for a disk galaxy. 

\subsection{Additional Radio Data}

     Radio observations of the \hi emission line were carried out with the 
DRAO Synthesis Telescope, covering a $2\deg$ field over the velocity range 
$-340 < v_{\rm LSR} < +500$ \kms. The resolution was $69\arcsec \times 
60\arcsec$ NS by EW and 4 \kms\, in velocity. The resulting sensitivity in 
four twelve--hour tracks was 13.3~mJy/beam in each velocity channel.  \hi 
emission from Cepheus~1 extends at least over the $v_{\rm LSR}$ range $+7$ 
to $+133$~km~s$^{-1}$.  The low--velocity emission is somewhat confused 
with \hi emission from the Galaxy, so the total velocity extent may still 
be underestimated. However, the systemic (LSR) velocity of the galaxy is 
well defined by the observed kinematics described below to be $65\pm 
1$~km~s$^{-1}$. Using the positive--velocity flank of the \hi emission 
profile and the systemic velocity allows determination of the linewidths at 
20\% and 50\% of the peak intensity, W$_{20}$~=~124~km~s$^{-1}$ and 
W$_{50}$~=~90~km~s$^{-1}$. An image of the integrated \hi column depth is 
shown in Figure~3; the corresponding velocity field is shown in Figure~4. 
Figure~3 shows a high--brightness gaseous disk with a diameter of about 
$7\arcmin$ in which column densities between $5\times 10^{20}$ to $10^{21}$ 
cm$^{-2}$ are observed. The \hi diameter defined by an azimuthally averaged 
face--on surface density of 1~M$_\odot$ pc$^{-2}$ is $11\farcm7$ and the 
integrated \hi flux is 136~Jy~\kms. The small ellipticity of the 
high--brightness \hi disk is consistent with inclinations of less than 
about $35\deg$.  Over this region, the kinematics are well described by a 
rotating disk system with constant inclination and a position angle of the 
receding line of nodes of about $-50\deg$. The kinematics indicate a spiral 
galaxy. The rotation velocity in the inner disk rises relatively slowly to 
a maximum value which is closely linked to the assumed inclination, but 
this parameter is difficult to determine, from the current data, for such a 
closely face--on system. If the inclination is as large as $35\deg$ the 
rotation velocity would only be about 60~km~s$^{-1}$, but for an 
inclination of 20$\deg$, it would be about 100~km~s$^{-1}$, and it would be 
even larger for a smaller inclination.

     Beyond the bright, inner \hi disk, there is a discontinuous change in 
the kinematics. An offset in the iso--velocity contours of some 
20~km~s$^{-1}$ is observed, in the sense of a shift to higher apparent 
rotation velocity. For this reason, the most extreme velocities are found 
in the outer--disk \hi arms that extend NW and S of the galaxy center. The 
outer--disk kinematics can be modeled by an increasing inclination and 
shifting of the line of nodes, but this solution is poorly constrained by 
the limited position angles at which outer disk gas is detected. A less 
likely possibility, although one which can not yet be ruled out, is that 
the inclination stays fixed while the rotation velocity continues to climb. 
If this were the case, the rotation velocity would climb to 100 \kms\, or 
190~km~s$^{-1}$, for inclinations of 35$\deg$ or 20$\deg$, respectively.

     The observed parameters of Cepheus 1 are summarized in Table 1.\\

\section{Physical Properties of Cepheus 1}

     Having established the presence of Cepheus~1, we considered its 
distance and physical properties.  Random and systematic motions of nearby 
galaxies can be at least as large as those expected from the Hubble flow, 
rendering the cosmological velocity/distance relation of little use for 
systems nearer than about 10 Mpc.  A preliminary distance estimate follows 
from the tendency of galaxies to cluster on all scales.  A galaxy is not 
commonly more remote from any neighbor than, say, 1 Mpc.  The case of 
NGC\,6946, one of the 3 or 4 nearest giant Sc spiral galaxies, is 
interesting in this regard.  It occupies an apparently lonely position (at 
$l,b = 95\fdg7,11\fdg7$) in the Local Void (\cite{tul87}), accompanied by 
its dwarf companion UGC\,11583 at a projected distance of $38'$, 
corresponding approximately to the linear distance of the Large Magellanic 
Cloud from our own Galaxy.  The distance of NGC\,6946 has recently been 
determined (\cite{sha97}) as $6.4\pm0.4$ Mpc, from photometry of resolved 
stars in it and its irregular dwarf companion, UGC\,11583, using the value 
A$_{\rm V}=1.93$ to correct for the foreground extinction.  The 
heliocentric velocity of NGC\,6946 is $+51\pm4$ \kms, corresponding to 
$v_{\rm GSR} = 280$ \kms. 

     In view of its angular and, especially, kinematic, proximity to 
NGC\,6946, it seems altogether plausible that Cepheus~1 is at a similar 
distance.  Cepheus~1 is situated south of NGC\,6946 some $3\fdg9$, 
corresponding at 6 Mpc to a projected linear distance of 410 kpc, which is 
about half the distance separating the Milky Way from our own nearest large 
neighbor, the Andromeda galaxy.  The kinematic separation between 
NGC\,6946 and the new galaxy is only 2 \kms\, in the $v_{\rm GSR}$ frame.  
Recently the Dwingeloo Obscured Galaxies Survey (\cite{hen98,riv98}) 
detected 21--cm signatures from three candidate galaxies within about 
$10\deg$ of NGC\,6946 and each with a recessional velocity $v_{\rm LSR} < 
250$ \kms.  If, as seems likely, these candidates are confirmed in optical 
and radio synthesis data to be dwarf galaxies, then NGC\,6946 and its 
accompanying systems will constitute a small group. 

     The distance of 6 Mpc, suggested as a working hypothesis, can be 
tested in several ways.  The APO H$\alpha$ image shows \hii regions 
scattered over the entire field; a deeper image of a larger field would 
plausibly reveal more such regions.  If we adapt $6'$ as the angular 
dimension and 6 Mpc as the distance, then the projected linear size of the 
\hii region distribution of Cepheus~1 is 10.5 kpc.  The linear size 
following from the $11\farcm7$ \hi dimension is 21 kpc; the total \hi gas 
mass is $M_{\rm HI} = 1.1\times 10^9$ M$_\odot$.  Cepheus~1 is thus 
comparable to the nearby spiral M33 (\cite{cor89,gui81}) in terms of the 
scale length of the stellar continuum (1.7 kpc for M33), the spatial extent 
of the \hii region distribution (15 kpc diameter for M33), and the total 
\hi mass ($1\times 10^9$ M$_\odot$).  The \hi diameter of Cepheus 1 (21 
kpc) is larger than that of M33 (15 kpc when defined in the same way).  
The disk of M33 is somewhat bluer (B--V=0.46) than that of Cepheus~1, and 
intrinsically brighter, by a factor 6 in total brightness and a factor 3 or 
4 in central surface brightness.  Of the 19 LSB galaxies in the sample of 
de Blok (1997), 7 have an \hi size smaller than that of Cepheus 1, and 12 
have a larger size.  Of the 48 HSB galaxies in Broeils's (1992) ``Sample 
1", 13 have an \hi size smaller than Cepheus 1, and 35 have a larger \hi 
size. Galaxies with an \hi diameter of about 20 kpc are typically 
classified as Sc or Sd.  Comparisons with the properties of other nearby 
LSB systems (see \cite{kar96}), in particular with those of nearest member 
of that class, NGC\,247 (\cite{car85}), will be made when additional data 
are available.  We note that the \hi diameter of Cepheus 1 is about as 
large as that measured for the Milky Way to the distance at which the \hi 
surface density of our own Galaxy begins to rapidly fall. 

     We determined the integrated magnitudes of Cepheus 1 from the APO 
data.  The apparent magnitude in the R band, integrating the disk profile 
to infinity, is in the range $m_{\rm R}=13.1$ to $13.5$.  Correcting for 
foreground extinction this implies $m_{\rm R}=11.5$ to $11.9$.  The 
absolute magnitudes for $D=6$ Mpc are in the range $M_{\rm R}=-17.4$ to 
$-17.0$ and $M_{\rm B}=-16.5$ to $-16.1$. The faintness of its disk is 
consistent with Cepheus 1 being an LSB galaxy, albeit a rather bright 
example of these systems.  The cautions expressed in Section 2.2 apply here 
as well, i.e. the optical disk parameters are not yet well constrained, 
which could affect the integrated optical magnitudes, and the actual 
central surface brightness estimate, which could be lower if there is in 
fact a faint bar present.  The uncertainties stem from the range of 
possible disk scale lengths.  No correction has yet been made for internal 
extinction within that disk, which may be low, if the dust--to--gas ratio 
is low; the \hi column with a normal dust--to--gas ratio would give 
$\tau_{\rm V} = 0.2$ -- 0.4.  Assuming an internal extinction in the B band 
of A$_{\rm B} = 0.3$ mag, suggests an absolute extinction--corrected B 
magnitude of about $-16.6$ mag.  The resulting ratio M$_{\rm {HI}}$/${\rm 
L}_{\rm B}$ is 1.6 M$_\odot$/L$_\odot$, for a value M$_{\rm B}=5.48$ for 
the Sun. This is a very high ratio of gas mass to optical luminosity for 
normal high--surface--brightness galaxies, and even moderately high for the 
class of LSB systems (\cite{deb97}). 
                              
     A further check on the distance follows from the luminosities of the 
\hii regions. We derived the cumulative H$\alpha$ luminosity function of 
the \hii regions in Cepheus 1, and compared it with that for M33. Taking 
foreground extinction into account, the two luminosity functions overlap if 
the distance to Cepheus 1 is taken as 6 Mpc.  While this constraint is not 
very strong, given the possible variation in \hii--region luminosity 
functions, a distance very different than 6 Mpc seems unlikely.  We have 
not yet derived a Tully--Fisher distance to the new galaxy. In  view of the 
intrinsic scatter (\cite{zwa95}) in the T--F relationship for LSB systems, 
and in view of the uncertainties in the inclination, i.e. in $\Delta v$, we 
postpone this derivation until the more detailed data expected are 
available.  

     The optical spectra also support the LSB classification.  Notably, we 
only detected emission from doubly--ionized oxygen in the \hii regions, but 
not from singly--ionized nitrogen or sulphur.  For the brighter \hii 
region, we derived limits on the line ratios of [NII](6584\,\AA)/H$\alpha$ 
$< 0.10$ and [SII](6716\,\AA)/H$\alpha$ $<0.12$, and a high ratio of 
[OIII](5007\,\AA)/H$\beta$ $=2.8$. These line ratios are entirely 
consistent with typical LSB \hii regions (\cite{imp97,bot97}), but not with 
the inner \hii regions in normal spirals. The ratios indicate a low 
abundance of heavy elements, a characteristic property of LSB galaxies.  An 
attempt was made on the JCMT to detect CO(J=2--1) from the central 
direction of Cepheus~1.  That the attempt returned a null spectrum is not 
decisive, because spiral galaxies show a wide range of molecular 
morphologies relative to \hi morphologies, but the lack of detectable CO is 
consistent with the situation (\cite{deb97}) in all other LSB systems.  
Because of the proximity of Cepheus 1, a very deep CO observation would 
plausibly establish a very good limit to the molecular content of an LSB 
galaxy. We note also the sparsely populated environment, where the 
evolution of Cepheus 1 proceeds with evidently little external influence on 
the internal physics.

     In view of the various uncertainties, our distance determination 
should be regarded as preliminary, although the suggested association with 
NGC\,6946 at about 6 Mpc is supported by all the data at our disposal.  
The evidence points toward the morphology of a low--surface--brightness 
spiral galaxy: the faint disk surface brightness, the scattered \hii 
regions, the substantial \hi columns, the relatively strong and regular \hi 
rotation signature, the weak molecular level, the optical color, the low 
heavy--metal abundance, and the relative isolation in the Local Void.  
Cepheus 1 is one of the nearest few LSB spirals and the nearest seen at low 
inclination.  It is a galaxy of moderate \hi size, and a rather extreme 
M$_{\rm {HI}}$/L$_{\rm B}$ ratio.  The discovery of nearby galaxies extends 
the knowledge of the local luminosity and mass functions.  The discovery of 
a likely member of the LSB class offers opportunities for studies in 
detail.  Such opportunities are rare because luminous LSB galaxies are 
mostly found at large distances, but exploiting what opportunities there 
are is important in view of the contributions such systems may make to 
understanding the formation and evolution of galaxies.  By itself, Cepheus 
1 would not significantly distort the dynamics of the Local Group. But it 
will be interesting to establish if NGC\,6946 and Cepheus 1 together with 
the other gas--rich members which are being found (\cite{hen98,riv98}) 
nearby might, as a group, influence the gravitational torques in the local 
neighborhood (see \cite{pee94} and \cite{lyn81}).  A multi--wavelength 
observational campaign has been prepared which will more tightly constrain 
the distance to the newly--unveiled galaxy and provide details of its 
structure.  

\begin{acknowledgments}  We are grateful to M. Irwin for scanning the 
three--color POSS II data using the RGO APM machine; to E. Turner for 
rapidly providing access to the APO 3.5--meter telescope; to A. Gray and T. 
Landecker at the DRAO for facilitating \hi mapping using the Penticton 
array; and to R. Tilanus and F. Baas for the CO observation at the JCMT.
\end{acknowledgments}

\newpage
\begin{figure}[tbp] 
\vspace{.1cm} 
\caption{\hi spectrum of the galaxy Cepheus~1 extracted from the 
Leiden/Dwingeloo survey (\cite{har97}), toward $l,b=94\fdg5,8\fdg0$. The 
velocity is relative to the Local Standard of Rest, the intensity is in 
terms of brightness temperature. The lower--velocity wing of the galaxy's 
signature is buried under emission from our own Galaxy; thus from this 
detection spectrum alone it is not possible to determine the total velocity 
width of the galaxy's signature, nor the velocity centroid giving the 
systemic velocity.} \end{figure} 

\begin{figure}[tbp] 
\vspace{.1cm}
\caption{Color--coded composite of the I--, R--, and H$\alpha$ images of 
Cepheus~1 observed with the 3.5--m telescope of the Apache Point  
Observatory.  The field size is $4\farcm4 \times 6\farcm8$ EW by NS. The 
red, green, and blue intensities in the image are proportional to the 
surface brightness in the I--band, R--band, and continuum--subtracted 
H$\alpha$ exposures, respectively.  Note the bar--like concentration of 
stellar continuum emission peaking near $\alpha,\,\delta\,({\rm J2000}) = 
20^{\rm h}51^{\rm m}10\fs6,+56\deg53'25\farcs0$, and the \hii regions which 
are visible rather uniformly scattered over a large portion of the field.  
H$\alpha$ spectra were obtained for two \hii regions, E and W of the 
center.} 
\end{figure} 

\begin{figure}[tbp]
\vspace{.1cm}
\caption{Integrated \hi emission in Cepheus~1 as observed with the DRAO 
Synthesis Telescope. The linear pseudo--color scale extends over apparent 
column densities of 100 to 800 $\times 10^{18}$ cm$^{-2}$ observed with a 
beam size of $69\arcsec \times 60\arcsec$ NS by EW. The high--brightness 
inner disk appears nearly circular, suggesting an almost face--on 
orientation.  With a size of about $7\arcmin$, the inner disk extends over 
a comparable field to that depicted optically in Figure~2. Fainter 
arm--like features extend further to both the North and South for a total 
gaseous extent of some $14\arcmin$.  The observed \NH\, corresponds to a 
hydrogen mass of $1.1\times 10^{9}$ M$_\odot$, at a distance of 6 Mpc.}
\end{figure} 

\begin{figure}[tbp] 
\vspace{.1cm}
\caption{Velocity field of the \hi emission in Cepheus~1 as observed with 
the DRAO Synthesis Telescope. The linear pseudo--color scale extends 
between $v_{\rm LSR} = +10$ and +100 \kms.  Note the classic 
spider--diagram pattern within the inner disk, signifying the well--ordered 
rotation characteristic of a spiral galaxy. The outer disk shows a 
kinematic discontinuity of more than 20 \kms\, with respect to the adjacent 
inner disk velocities, suggesting either strong warping or a rising 
rotation curve in the outer disk. }
\end{figure}

\newpage
\begin{deluxetable}{lcr}
\tablecolumns{3}
\setlength{\tabcolsep}{1pt}
\scriptsize
\tablecaption{Properties of Cepheus 1\label{tbl-1}}
\tablewidth{0pt}
\tablehead{
\colhead{Parameter}      &\colhead{Value} & \colhead{Notes} } 
\startdata
Optical center: $\alpha,\delta$(J2000)& $20^{\rm h}51^{\rm m}10\fs6 
(\pm5''), +56^\circ53'25\farcs0(\pm5'')$ & \ \nl
Dynamical center: $\alpha,\delta$(J2000)& $20^{\rm h}51^{\rm m}10\fs8 
(\pm5''), +56^\circ53'26\farcs4(\pm5'')$ & \ \nl 
Galactic coordinates: $l,b$  & $94\fdg375$, $+8\fdg010$ & \ \nl  
Distance & 6 Mpc & \tablenotemark{a} \nl
Optical axis ratio & $0.4\pm0.1$ & \tablenotemark{b} \nl
Foreground extinction, A$_{\rm V}$ & $<2.45$ mag & \tablenotemark{c} \nl
Adopted foreground extinction, A$_{\rm V}$ & $2.0\pm0.2$ mag & 
\tablenotemark{d} \nl
Internal extinction, A$_{\rm B}$ & $0.3\pm0.2$ mag & \tablenotemark{e} \nl
Central R surface brightness, $\mu_0^{\rm R}$ & $21.9\pm0.2$ 
mag~arcsec$^{-2}$ & \tablenotemark{b,f}  \nl
Central B surface brightness, $\mu_0^{\rm B}$ & $22.9\pm0.2$ 
mag~arcsec$^{-2}$ & \tablenotemark{b,f}  \nl
Absolute R magnitude, M$_{\rm R}$ & $-17.5\pm0.3$ mag & 
\tablenotemark{g}  \nl
Absolute B magnitude, M$_{\rm B}$ & $-16.6\pm0.3$ mag & 
\tablenotemark{g}  \nl
Total blue luminosity, L$_{\rm B}$ & $6.8\pm2.2 \times 10^8$ L$_{\rm 
B}\odot$ &
\tablenotemark{g,h} \nl
Stellar disk scalelength, $s_R$ & 60 arcsec, 1.8 kpc & \tablenotemark{b}  
\nl
Total \hi flux, F$_{\rm HI}$ & $136\pm10$ Jy km s$^{-1}$ & \ \nl
Total \hi mass, M$_{\rm HI}$ & $1.1\pm0.1 \times 10^9$ M$_\odot$ & \ \nl
Systemic velocity, $v_{\rm HEL}$ & $+58\pm3$ km s$^{-1}$ & \ \nl
Line width at 50\% intensity, W$_{50}$ & $90\pm5$ \kms & \ \nl
Line width at 20\% intensity, W$_{20}$ & $124\pm5$ \kms & \ \nl
Inclination & $28\pm8^\circ$ & \  \nl
Position angle & $-50\pm3^\circ$ &\ \nl
Diameter of \hi disk, D$_{\rm HI}$ & 11\farcm7, 21 kpc & \tablenotemark{i} 
\nl
\enddata
\tablenotetext{a}{Assuming association with NGC~6946 as discussed
in the text.}
\tablenotetext{b}{May be strongly affected by the presence of a bar.} 
\tablenotetext{c}{Based on Balmer decrements of two \hii regions.} 
\tablenotetext{d}{Based on Galactic N$_{\rm HI}$.}
\tablenotetext{e}{Based on Cepheus 1 N$_{\rm HI}$.}
\tablenotetext{f}{Corrected for foreground extinction.}
\tablenotetext{g}{Corrected for foreground and internal extinction.} 
\tablenotetext{h}{Assuming M$_{\rm B}\odot$ = 5.48.}
\tablenotetext{i}{Measured at a mass surface density of 1 M$_\odot$ pc 
$^{-2}$.}
\end{deluxetable}

\end{document}